\documentclass[10pt,letterpaper,twocolumn]{article} 
\usepackage{ol2}

\usepackage{amsmath}

\begin{document}

\twocolumn[ 

\title{Photonic analogue of {\it Zitterbewegung} in binary waveguide arrays}


\author{S. Longhi}
\address{Dipartimento di Fisica, Politecnico di Milano, Piazza L. da Vinci 32, I-20133 Milano, Italy}
\begin{abstract}
An optical analogue of {\it Zitterbewegung} (ZB), i.e. of the
trembling motion of Dirac electrons caused by the interference
between positive and negative energy states, is proposed for spatial
beam propagation in binary waveguide arrays. In this optical system
ZB is simply observable as a quiver spatial oscillatory motion of
the beam center of mass around its mean trajectory.
\end{abstract}
\ocis{230.7370, 000.2658, 050.5298}


] 

\noindent

Zitterbewegung (ZB) refers to the rapid trembling motion of a free
Dirac electron caused by the interference between positive and
negative energy state components \cite{ZB1,ZB3}. However, the
amplitude of ZB oscillations turns out to be extremely small, of the
order of the Compton wavelength, which defines the limit of electron
localization. Therefore, its direct observation is unlikely, though
observable consequences of ZB have been found in the response of
electrons to external fields \cite{ZB1}. Similarly to other effects,
like the Klein paradox \cite{ZB3}, ZB has been for long time
regarded as a relativistic effect rooted in the Dirac equation.
However, several authors have recently shown that ZB is not unique
to Dirac electrons, rather it is a generic feature of wave packet
dynamics in spinor systems with certain linear dispersion relations,
such as those exhibiting the so-called Dirac points (DP) that
describe massless fermions
\cite{Lamata07,Cserti06,Rusin07,Vaishnav08,Haldane,
Sepkhanov07,Peleg07,Peleg08,Zhang08,Wang09,Wang09b}. In
condensed-matter and matter-wave physics, trapped ions
\cite{Lamata07}, graphene \cite{Cserti06,Rusin07}, and ultracold
neutral atoms \cite{Vaishnav08} have been proposed as candidate
systems for a direct observation of ZB. Similarly, in the optical
context it was shown that DP can occur in two-dimensional photonic
crystals \cite{Haldane,Sepkhanov07,Peleg07,Peleg08,Zhang08} or in
negative-zero-positive index metamaterials \cite{Wang09}. Photonic
analogues of ZB have been recently proposed in
Refs.\cite{Zhang08,Wang09b}, in which the observation of ZB requires
time-resolved measurements of pulse transmission through photonic
crystal or metamaterial slabs of different thickness. In this Letter
we propose a simpler optical system, consisting of a binary
waveguide array, which enables an easy visualization {\it in space}
of photonic ZB. The ability of mapping typical ultrafast phenomena
occurring in the matter as spatial propagation of light waves in
coupled waveguide structures has been successfully demonstrated in
several experiments (see, e.g.,
\cite{Trompeter06,Longhi09,DellaValle09,Dreisow09}), and our
proposal should thus greatly facilitate the way toward a first
observation of ZB. \par Let us consider propagation of monochromatic
light waves at wavelength $\lambda$ in a one-dimensional optical
lattice, which is described by the following scalar equation for the
electric field envelope $E(x,z)$
\begin{equation}
i \partial_z E  = -[\lambda/(4 \pi n_s)] \partial^2_xE + (2
\pi/\lambda)[n_s-n(x)] E,
 \end{equation}
\begin{figure}[htb]
\centerline{\includegraphics[width=8.2cm]{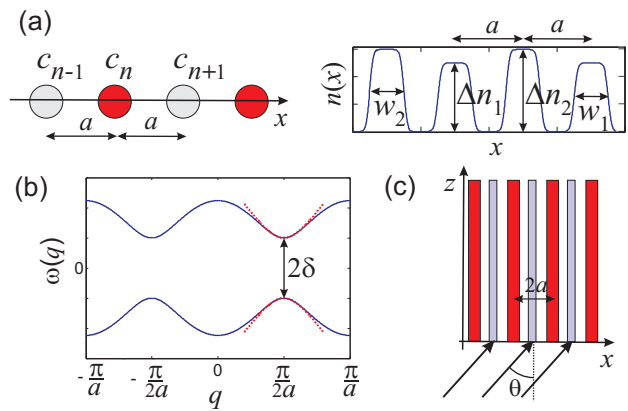}} \caption{
(Color online) (a) Schematic of a one-dimensional binary array
(left) and refractive index profile (right). (b) Dispersion curves
of the first two minibands of a tight-binding binary array (solid
curves), and corresponding dispersion curves of the Dirac equation
(4) (dotted curves). (c) Broad-beam excitation geometry of the array
at a tilting angle $\theta$.}
\end{figure}
where $n_s$ is the substrate refractive index and $n(x)$ is the
refractive index profile of the lattice. To realize a photonic
analogue of ZB,  a two-band model with a dynamics described by a
two-component spinor wave function is needed. Such a model can be
realized by either a singly-periodic lattice with a shallow
sinusoidal refractive index profile, where the two components of the
spinor wave functions  are related by simple linear transformation
to the amplitudes of counterpropagating waves in the lattice (see,
for instance, \cite{Feng03}), or to a tight-binding binary
superlattice composed by two interleaved sublattices A and B [see
Fig.1(a)], where the spinor wave functions correspond to the
occupation amplitudes in the two sublattices. Here we study ZB in
the latter structure and assume a symmetric intersite coupling,
which corresponds to the experimental conditions of
Ref.\cite{Dreisow09}; the analysis could be extended, if needed, for
the more general case of non-symmetric intersite couplings
\cite{Kivshar02}. In the tight-binding approximation, light
transport in the binary lattice is described by coupled-mode
equations for the modal field amplitudes $c_n$ in the various
waveguides \cite{Dreisow09,Kivshar02,Morandotti04}
\begin{equation}
i (dc_n/dz)=-\sigma (c_{n+1}+c_{n-1})+(-1)^n \delta c_n,
\end{equation}
where $2 \delta$ and $\sigma$ are the propagation constant mismatch
and the coupling rate between two adjacent waveguides of the array,
respectively. The tight-binding model (2) supports two minibands,
whose dispersion curves are readily obtained by the plane-wave
Ansatz $c_n(q) \sim \exp(iqna-i\omega z)$ and read \cite{Kivshar02}
\begin{equation}
\omega(q)= \pm \sqrt{\delta^2+4 \sigma^2 \cos^2 (qa)}
\end{equation}
[see Fig.1(b)]. Let us assume that the array is excited by a broad
beam (e.g. Gaussian shaped) incident onto the array at an angle
close to the Bragg angle $\theta=\lambda/(4n_sa)$, i.e. $E(x,0)=G(x)
\exp(2 \pi i \theta  n_s x / \lambda)$, where $a$ is the spacing
between adjacent waveguides and $G(x)$ varies slowly on the spatial
scale $\sim a $ [see Fig.1(c)]. At such an incident angle, the modes
in adjacent waveguides are excited with a nearly equal amplitude but
with a phase difference of $\pi /2$. After setting
 $c_{2n}(z)=(-1)^n\psi_1(n,z)$ and $c_{2n-1}=-i(-1)^n \psi_2(n,z)$,
for broad beam excitation the amplitudes $\psi_1$ and $\psi_2$ vary
slowly with $n$, and one can thus write $\psi_{1,2}(n \pm
1,z)=\psi_{1,2}(n ,z) \pm (\partial \psi_{1,2}/\partial n)$ and
consider $n \equiv \xi$ as a continuous variable rather than as an
integer index. At the input plane $z=0$, the amplitudes
$\psi_1(\xi,0)$ and $\psi_2(\xi,0)$ are proportional to $G(2na)$ and
$G(2na-a) \simeq G(2na)$, respectively, so that one can assume
$\psi_1(\xi,0) \simeq \psi_2(\xi,0)$ as an initial condition. Under
such assumptions, from Eqs.(2) it readily follows that the
two-component spinor $\psi(\xi,z)=(\psi_1,\psi_2)^T$ satisfies the
one-dimensional Dirac equation
\begin{equation}
i \partial_z \psi+i \sigma \alpha \partial_{\xi} \psi-\delta \beta
\psi=0,
\end{equation}
where
\begin{equation}
\alpha=\left(
\begin{array}{cc}
0 & 1 \\
1 & 0
\end{array}
\right) \; , \; \beta=\left(
\begin{array}{cc}
1 & 0 \\
0 & -1
\end{array}
\right).
\end{equation}
Note that $\alpha$ and $\beta$ coincide with the  $\sigma_x$ and
$\sigma_z$ Pauli matrices, respectively. Assuming the normalization
condition $\int d \xi (|\psi_1|^2+|\psi_2|^2)=1$, after the formal
change $\sigma \rightarrow c$, $\delta \rightarrow m c^2 / \hbar$, $
\xi \rightarrow x$ and $z \rightarrow t$, Eq.(4) corresponds to the
one-dimensional Dirac equation for an electron of mass $m$ in
absence of external fields \cite{ZB3}. The {\it temporal} evolution
of the spinor wave function $\psi$ for the Dirac electron is
therefore mapped into the {\it spatial} evolution of the modal
amplitudes $\psi_1$ and $\psi_2$ in the two sublattices A and B. The
energy-momentum dispersion relation $\hbar \omega(k)$ of the Dirac
equation (4), obtained by making the Ansatz $\psi \sim \exp(ik \xi-i
\omega z)$ in Eq.(4), is composed by the two branches $\omega(k)=
\pm \epsilon(k)$, corresponding to positive and negative energy
states of the relativistic free electron, where
$\epsilon(k)=\sqrt{\delta^2+\sigma^2 k^2}$. Such branches reproduces
the two minibands of the binary array [see Eq.(3)] near the boundary
of the Brillouin zone [see Fig.1(b)], where $k = 2 a q - \pi$. ZB of
the Dirac electron corresponds to a rapid oscillation of the average
position $\langle \xi \rangle(z)=\int d \xi \; \xi
(|\psi_1|^2+|\psi_2|^2)$ around the classical trajectory. The usual
method of understanding ZB in the framework of the Dirac equation is
to derive equations of motion for the Heisenberg operators, and show
that they oscillate in time \cite{ZB1,ZB3}. We instead work directly
in the Schr\"{o}dinger picture and consider wave packet evolution in
momentum space, namely we set $\psi_{1,2}(\xi,z)=\int dk
\hat{\psi}_{1,2}(k,z) \exp(ik \xi)$ and calculate the spectra
$\hat{\psi}_{1,2}(k,z)$ by solving Eq.(4) in momentum space. One
then obtains $\hat{\psi}_{1,2}(k,z)=\hat{G}(k)[\cos( \epsilon z) \mp
i( \pm \sigma k+\delta) \sin(\epsilon z) / \epsilon ]$, where
$\hat{G}(k)=(1/2\pi) \int d\xi G(2 a \xi) \exp(-ik \xi)$ is the
Fourier spectrum of the exciting broad input beam. The average
position $\langle \xi \rangle$ is then calculated as $\langle \xi
\rangle(z)=2 \pi i \int dk [\hat{\psi}_1^*(k)
\partial_k \hat{\psi}_1(k)+ \hat{\psi}_2^*(k) \partial_k
\hat{\psi}_2(k)]$, which yields after some algebra
\begin{eqnarray}
\langle \xi \rangle(z)  & = &  \langle \xi \rangle(0)+ 4 \pi
\sigma^3 z \int dk (k / \epsilon)^2 |\hat{G}(k)|^2+ \nonumber \\
& + &2 \pi \sigma \delta^2 \int dk (1/ \epsilon^3) \sin (2 \epsilon
z ) |\hat{G}(k)|^2
\end{eqnarray}
The last oscillatory term in Eq.(6) is ZB, superimposed to the
straight trajectory defined by the first two terms on the right hand
side of Eq.(6). For $\hat{G}(k)$ spectrally narrow at around $k=0$,
the frequency of ZB is equal to $2 \epsilon(k=0)=2 \delta$; spectral
broadening of $\hat{G}(k)$ is responsible for damping of ZB. It
should be noted that $\langle \xi \rangle$, calculated as the
average position for the Dirac equation (4), basically reproduces
the evolution of the beam center of mass $\langle n \rangle$ in real
space, defined as $\langle n \rangle =(\sum_n n|c_n|^2) / \sum_n
|c_n|^2$. In fact, it is straightforward to show that
\begin{equation}
\langle n \rangle=2 \langle \xi \rangle +\frac{1}{2}-4 \pi \delta
\sigma \int dk (k/ \epsilon^2) |\hat{G}(k)|^2 \sin^2(\epsilon z).
\end{equation}
\begin{figure}[htb]
\centerline{\includegraphics[width=8.2cm]{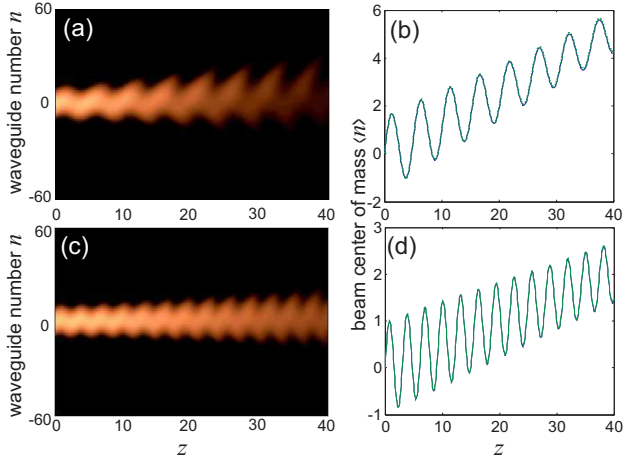}} \caption{
(Color online) (a) Evolution of $|c_n(z)|^2$ in a binary array
excited by a broad Gaussian beam for $\sigma=1$ and $\delta=0.6$,
and (b) corresponding behavior of beam trajectory $\langle n \rangle
(z)$ (solid curve) as obtained by numerical analysis of Eqs.(2). The
dotted curve in (b), almost overlapped with the solid curve,
reproduces the behavior of $2 \langle \xi \rangle (z)+1/2$, where
$\langle \xi \rangle (z)$ is calculated according to Eq.(5). (c) and
(d): same as (a) and (b), but for $\delta=1$.}
\end{figure}
The last term in Eq.(7) is usually negligible for a spectrum
$\hat{G}(k)$ narrow at around $k=0$ (see the examples to be
discussed below), so that one has $\langle n  \rangle \simeq 2
\langle \xi \rangle +1/2$. As an example, Fig.2 shows typical
evolutions of field intensities $|c_n(z)|^2$ [Figs.2(a) and (c)],
and corresponding behavior of beam center of mass $\langle n \rangle
(z)$ [Figs.2(b) and (d), solid curves], for $\sigma=1$ and for two
values of detuning $\delta$. In both cases, the array has been
excited by a broad Gaussian beam $G(x)= \exp[-(x/w_0)^2]$ with
$w_0/a=12$, and the results have been obtained by numerical
integration of coupled-mode equations (2) with initial conditions
$c_n(0)=i^n \exp[-(na/w_0)^2]$. A clear trembling motion of the
beam, corresponding to ZB, is observed, with an oscillation
frequency (amplitude) which increases (decreases) as $\delta$
increases, according to Eq.(6) [compare Fig.2(b) and Fig.2(d)]. A
similar trembling motion would be observed for a binary lattice with
asymmetric intesrite coupling \cite{Kivshar02}. Damping of ZB
oscillations, which arises from the broadening of the spectrum
$\hat{G}(k)$, is also visible in Figs.2(b) and (d); note also that
the averaged beam path has a nonvanishing drift velocity which
arises from the second term on the right hand side of Eq.(6).\\
\begin{figure}[htb]
\centerline{\includegraphics[width=8.2cm]{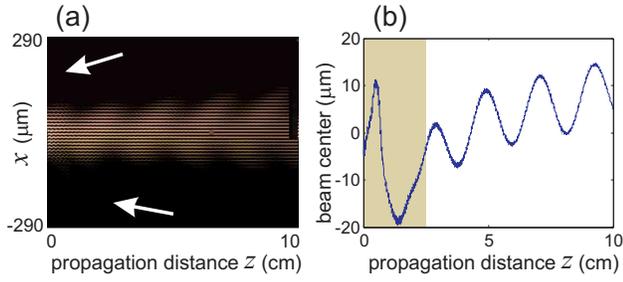}} \caption{
(Color online) (a) Beam propagation (snapshot of $|E(x,z)|^2$) in a
binary array, as obtained by numerical analysis of Eq.(1), for broad
Gaussian beam excitation tilted at the Bragg angle; parameter values
are given in the text. (b) Corresponding behavior of $\langle x
\rangle (z)$.}
\end{figure}
We finally checked the correctness of the analysis, based on the
tight-binding model (2), and the feasibility of an experimental
observation of ZB by numerical simulations of the wave equation (1)
for parameter values that typically apply to binary arrays realized
in fused silica by femtosecond laser writing \cite{Dreisow09}. A
typical numerical result is shown in Fig.3 for parameter values
$\lambda=633 \; {\rm nm}$, $n_s=1.42$, $a=10 \; \mu$m, $w_1=w_2=3.5
\; \mu$m, $\Delta n_1=0.003$, $\Delta n_2=0.00297$ and for an input
Gaussian beam $E(x,0)$ of spot size $w_0=80 \; \mu$m, tilted at the
Bragg angle $\theta \simeq 0.64^{\rm o}$. The trembling motion of
the beam as it propagates along the 10-cm-long array is clearly
visible, and should be easily observed by microscope fluorescence
imaging \cite{Dreisow09}. The beam trajectory in Fig.3(b) has been
computed as $\langle x \rangle (z) =\int dx \; x |E|^2$ / $\int dx
|E|^2$, where the integration interval is limited by the size of the
numerical domain. Note that the first oscillations of $\langle x
\rangle $ internal to the shaded area of Fig.3(b) do not correspond
to ZB, rather they arise because of an initial beam break up and
appearance of higher-order beams [as indicated by the arrows in
Fig.3(a)] belonging to higher-order bands of the array. Such
higher-order beams, however, refract at large angles and, after few
centimeter propagation, they are no
more overlapped with the main beam undergoing ZB.\\
In conclusion, a photonic analogue of the trembling motion of Dirac
electrons has been proposed for spatial beam propagation in binary
waveguide arrays. As compared to previous proposals
\cite{Zhang08,Wang09b}, the easy experimental visualization of beam
dynamics in waveguide arrays \cite{Dreisow09} should greatly
facilitate the way toward the first observation of ZB.

 Author E-mail address: longhi@fisi.polimi.it

\end{document}